\documentclass[conference]{IEEEtran}
\IEEEoverridecommandlockouts
\usepackage{cite}
\usepackage{amsmath,amssymb,amsfonts}
\usepackage{algorithmic}
\usepackage{graphicx}
\usepackage{textcomp}
\usepackage{textgreek}
\usepackage{listings}
\usepackage{xcolor}
\def\BibTeX{{\rm B\kern-.05em{\sc i\kern-.025em b}\kern-.08em
    T\kern-.1667em\lower.7ex\hbox{E}\kern-.125emX}}
\begin{document}

\title{ Variable Record Table: A Unified Hardware-Assisted Framework for Runtime Security}
\author{\IEEEauthorblockN{Suraj Kumar Sah}
\IEEEauthorblockA{\textit{Department of Computer Science and Engineering} \\
\textit{Kathmandu University}\\
Dhulikhel, Nepal \\
surajsah2053@gmail.com}
\and
\IEEEauthorblockN{Love Kumar Sah}
\IEEEauthorblockA{\textit{Department of Electrical and Computer Engineering} \\
\textit{Western New England University}\\
Springfield, MA, USA \\
love.sah@wne.edu}
}

\maketitle
\begin{abstract}
\label{sec:abs}
Modern computing systems face security threats, including memory corruption attacks, speculative execution vulnerabilities, and control-flow hijacking. Although existing solutions address these threats individually, they frequently introduce performance overhead and leave security gaps. This paper presents a Variable Record Table (VRT) with a unified hardware-assisted framework that simultaneously enforces spatial memory safety against buffer overflows, back-edge control-flow integrity (CFI), and speculative execution attack detection. The VRT dynamically constructs a protection table by instrumenting runtime instructions to extract memory addresses, bounds metadata, and control-flow signatures. Our evaluation across MiBench and SPEC benchmarks shows that VRT successfully detects all attack variants tested with zero additional instruction overhead. Furthermore, it maintains memory requirements below 25KB (for 512 entries) and maintains area / power overhead under 8\% and 11.65 μW, respectively. By consolidating three essential security mechanisms into a single hardware structure, VRT provides comprehensive protection while minimizing performance impact.
\end{abstract}


\section{Introduction}
\label{sec:intro}
\vspace*{-1ex}
Memory safety violations are among the most critical vulnerabilities in modern systems, with buffer overflows~\cite{stack-smashing}, control-flow hijacking~\cite{abadi2005}, and speculative execution attacks~\cite{kocher2018spectre} being the three primary threat classes. Despite the availability of various mitigation techniques, existing solutions have three fundamental limitations: (1) narrow protection scope (defending against only one attack class), (2) significant performance overhead from software mediation, and (3) security gaps between disjoint mechanisms.

The primary challenge in memory protection involves a trilemma between completeness, performance, and security. Current solutions force designers to choose between different approaches: using multiple-point solutions for comprehensive coverage (completeness), relying solely on hardware mechanisms for improved performance, or striving to close all vulnerability gaps (security). For example, while tagged memory architectures~\cite{cheri2015} offer strong spatial memory safety, they do not adequately address control-flow integrity or threats from speculative execution. Conversely, control-flow integrity mechanisms~\cite{abadi2005} focus only on validating branches, leaving them susceptible to memory corruption attacks. Software-based approaches, such as bounds checking, often attract significant overheads, typically exceeding 30\%\cite{szekeres2013}. Even hardware-assisted solutions like Intel's Control-flow Enforcement Technology~\cite{intel2016} protect against only certain types of vulnerabilities. Recent research~\cite{evans2015} has shown that sophisticated attacks can exploit the gaps between these isolated protections, underscoring the necessity for a unified solution. This solution should effectively address memory safety, control-flow integrity, and speculative execution threats, all while maintaining performance.

To address the limitations of existing security methods, we extend the \textit{Variable Record Table (VRT)}~\cite{self1,self2,self3,self4,self5}, a comprehensive hardware framework designed to enforce memory safety, control-flow integrity, and protection against speculative execution. VRT achieves this through three key innovations. First, it features a novel metadata architecture that captures variable bounds, control-flow signatures, and speculative access patterns within a single hardware structure, eliminating the need for separate protection mechanisms. Second, our design incorporates lightweight instrumentation of runtime instructions, enabling the dynamic construction of protection policies without requiring software intervention. This approach maintains zero additional instruction overhead. Third, VRT implements parallel security checks through a dedicated pipeline stage that performs bounds verification (for spatial safety), return address validation (back-edge CFI), and speculative access tagging in a single clock cycle. This unified approach provides three fundamental advantages over existing solutions: (1) \textit{comprehensive protection} against all three classes of attacks through shared metadata, (2) \textit{no impact on performance}, and (3) \textit{practical hardware costs}, with only a 1.98\% increase in area overhead. By consolidating traditionally separate security mechanisms into a coherent architectural framework, VRT provides protection against modern multi-vector attacks without compromising system performance.

The remainder of this paper is organized as follows. Section \ref{sec:lit} provides necessary background and existing approaches.  Section \ref{sec:architecture} presents the VRT architecture, including its metadata extraction mechanism, protection table design, and enforcement policies. Section \ref{sec:results} evaluates VRT's security coverage, performance impact, and hardware overhead through comprehensive experiments. Finally, Section \ref{sec:conclusion} discusses implications and future directions while concluding the paper.

\section{RUNTIME DEFENCE ARCHITECTURE}
\label{sec:architecture}

\begin{figure}[h]
\centering
\includegraphics[width=\columnwidth]{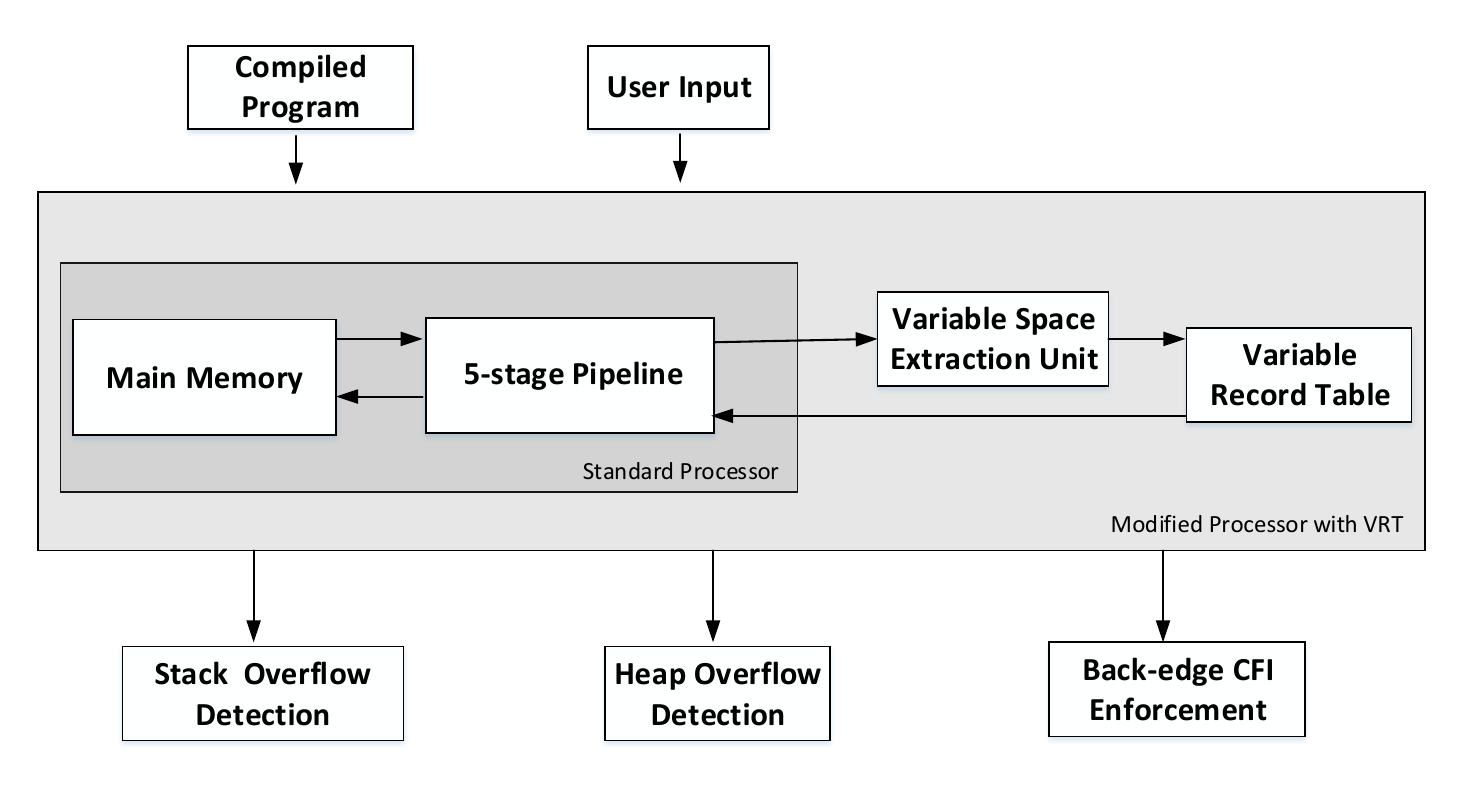}
\caption{Overall Architecture}
\label{fig:stack_memory}
\end{figure}

The architecture of the proposed system is shown in \mbox{Figure \ref{fig:stack_memory}}, where we augment the standard 5-stage pipeline processor with dedicated memory (VRT) to log runtime variable memory space information. This enhancement enables us to extract variable details in real-time from instructions interacting with the main memory, allowing us to verify their usage once control returns from a function.

\subsection{Variable Space Extraction Architecture}

\begin{figure}[h]
\centering
\includegraphics[width=\columnwidth]{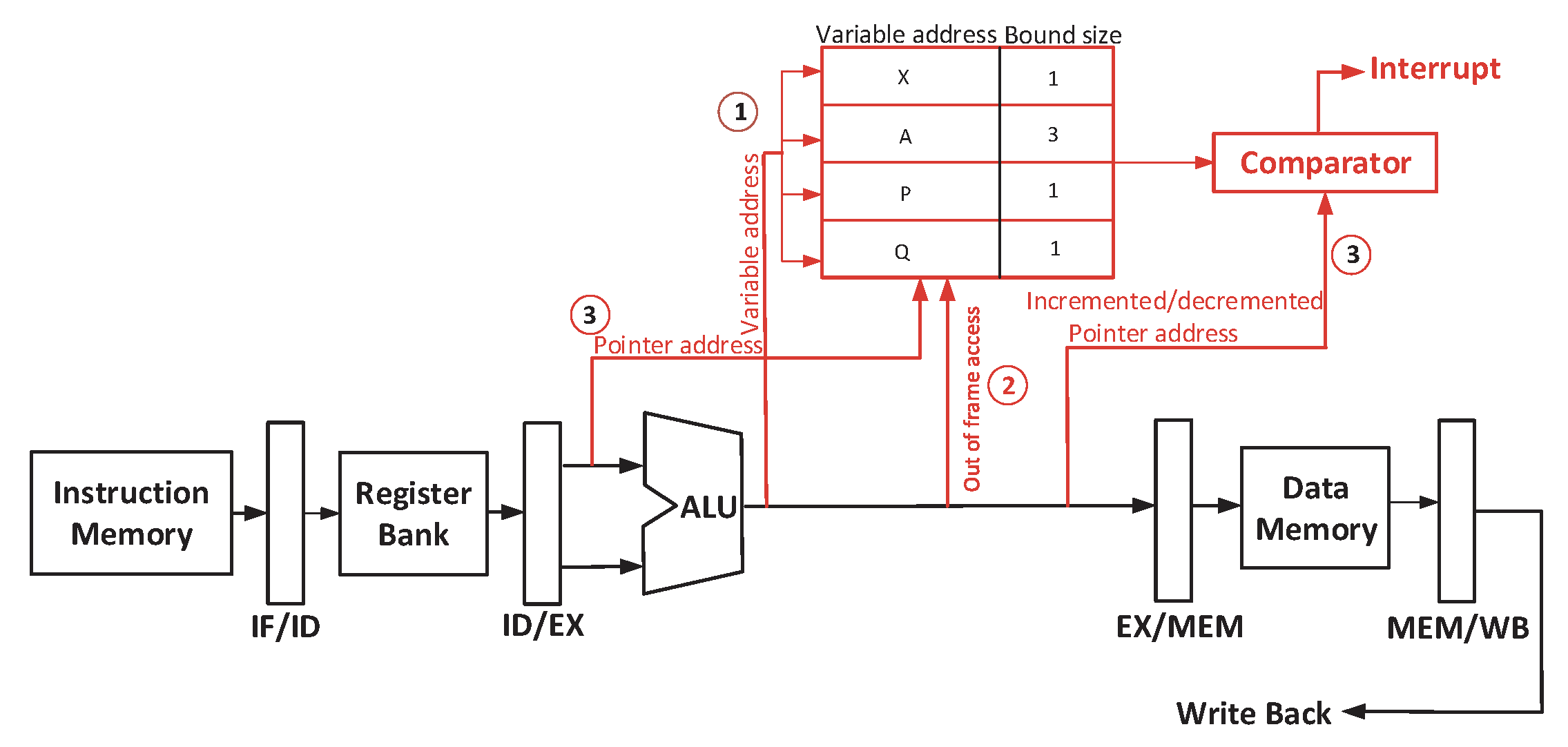}
\caption{Modified Pipeline for Runtime Variable Space Extraction}
\label{fig:micro-arch}
\end{figure}

Figure \ref{fig:micro-arch} illustrates the architecture for extracting base and bound information during runtime, specifically during the decode and execution stages. This runtime instrumentation specifically targets instructions that could generate a new address potentially associated with the frame pointer, which is stored in a special register as indicated in Table 1. 


Table \ref{tab:VRT} presents the layout of the VRT, which consists of three columns: the associated bit, the variable base address, and the bound value. Each entry in the VRT includes an associated bit, a 32-bit base address, and an 8-bit bound value, resulting in a total of 41 bits per entry. The allied bit differentiates entries for subsequent function calls. This table snippet displays six entries from the active function and two from the preceding function.

\subsection{Buffer Overflow and VRT} 


After populating the VRT with base and bound addresses of local variables, we can evaluate each array offset and pointer operation to identify potential invalid memory addresses in two representative scenarios. In this section, we will discuss both cases of illegal access.

\label{sec:VRT}
\begin{table}[h]
\begin{center}
\caption{Variable Record Table}
\label{tab:VRT}
\begin{tabular}{|c|c|c|}
\hline
Associated & Variable Address & Bound\\ \hline
\textbf{1} & 0X7FFF60 & 24 \\ \hline
\textbf{1} & 0X7FFF3B & 4 \\ \hline
\textbf{1} & 0X7FFF38 & 4 \\ \hline
\textbf{1} & 0X7FFF34 & 4 \\ \hline
\textbf{1} & 0X3FFF30 & 24 \\ \hline
\textbf{1} & 0X3FFE28 & 256 \\ \hline
0 & 0X7FFE70 & 24 \\ \hline
0 & 0X7FFE60 & 16 \\ \hline
\end{tabular}
\end{center}
\end{table}
\subsubsection{Constant variable index}


The first case involves direct access to an array using a constant index that exceeds the array's range, which can result in out-of-bounds access. If this operation is unchecked, it may corrupt data outside the allocated scope. In C code, this appears to be an attempt to access an array with an out-of-bounds index:

\begin{verbatim}
a[out_of_bound] = 'X';
\end{verbatim}

The corresponding assembly instruction for array access illustrates how the offset involved in the load instruction can lead to an address outside the valid address space of the variable stored in the VRT:

\begin{verbatim}
4002e0: lw $2,out_of_bound($30)
\end{verbatim}

\subsubsection{Loop operation on array or pointer variable} 

This issue often arises in buffer overflow scenarios, particularly with string library functions like `strcpy()` during loop operations. An unchecked increment of a pointer variable can result in addresses that exceed the allocated memory space:
\begin{verbatim}
char X[6];
char *ptr = X;
for(i=0; i<10 ;i++)
   ++ptr = '\0';
\end{verbatim}

Furthermore, this pointer increment operation can demonstrated using MIPS-like assembly instructions, where register \$2 serves the source and destination address. In out-of-bounds cases, \$2 may contain addresses that span multiple entries in the VRT, whereas valid operations will remain within a single VRT entry.

\begin{verbatim}
4002e0: lw $2,44($30) 
4002e8: addu $3,$0,$2
4002f0: sll $2,$3,0x2  
4002f8: lw $3,40($30)
400300: addu $2,$2,$3
\end{verbatim}


To mitigate these issues, the pipeline implementation adds VRT checks during the execution stage. When an out-of-bound access is detected during address generation, the operation is blocked before it can corrupt memory.

\subsection{Backward-edge CFI Enforement}
\begin{figure}[h]
\centering
\includegraphics[width=\columnwidth]{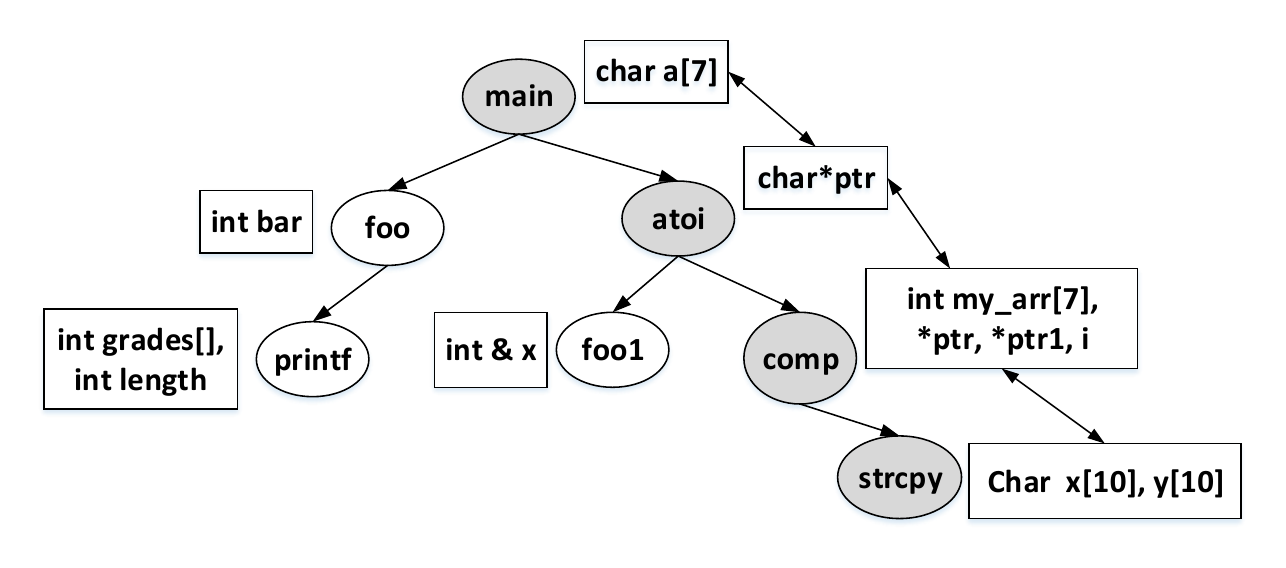}
\caption{A Control Sequence Graph  Example.}
\label{fig:Runniample}
\end{figure}

 In a Control Flow Graph (CFG), a backward edge indicates a transfer of control back to a preceding node within the graph. This occurs due to the `ret` instruction, which directs the control flow to the instruction immediately following a function call. Figure \ref{fig:Runniample} demonstrates how the control flow in a program initiates multiple function calls, each taking its unique execution path and managing a specific data set.
 
Backward-edge Control Flow Integrity (CFI) ensures that functions return to the correct location by verifying the return address stored in the stack frame. Acknowledging that this return address can be vulnerable to various attacks, we propose an additional validation method that uses the base addresses of stored variables. When a function returns, the program continues to use the same variable set. Our approach involves checking the first memory address accessed by load (lw) and store (sw) operations after the function returns. If this address matches one of the variables in our predetermined list, we consider the control flow path normal. In cases where the control flow may be compromised, we anticipate two distinct scenarios:

\begin{enumerate}
    \item The return lead to the beginning of an entirely different function: This is typically decoded by observing an instruction such as  \verb|subl $16, %esp|, where stack space is allocated for the new function call. Thus the return instruction anomalously precedes the instruction that creates stack space, indicating a potential compromise.


    \item The return leads to an arbitrary address: In this scenario, the addresses generated from load/store operations subsequent to a return instruction do not match any current variable in the variable record table. Therefore, after a return is executed, our system is tasked to verify these generated load/store addresses align with an entry for the expected returning function in the variable record table. This validation process is essential for detecting returns to unintended or malicious locations, thus maintaining the integrity of the program's control flow.
\end{enumerate}

\subsection{Defending Against Cache Probe Attacks Using VRT}

\begin{figure}
\centering
\includegraphics[width= 0.5\textwidth]{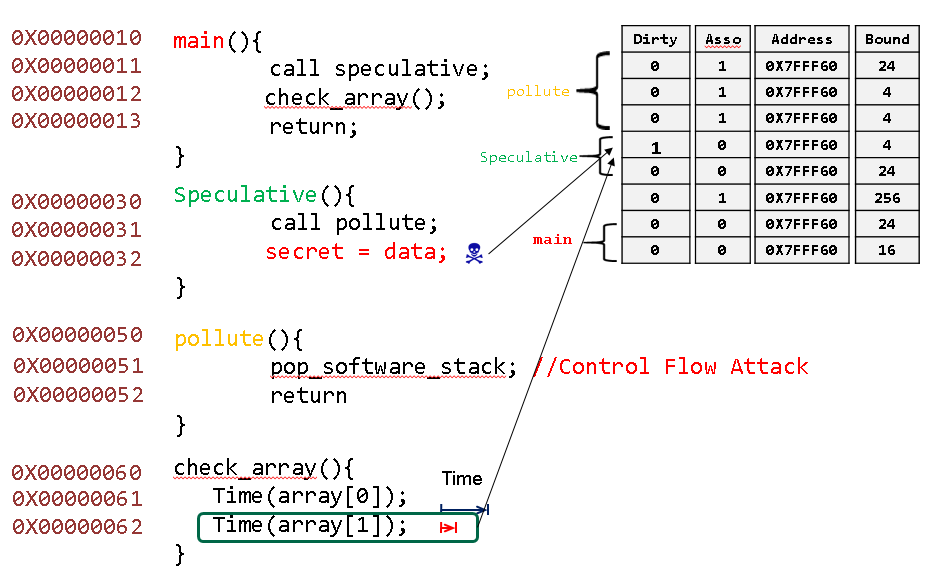}
\caption{VRT with dirty bit and cache probe  }
\label{fig:dirt}
\end{figure}

The VRT mechanism Figure \ref{fig:dirt} provides  protection against cache probe attacks by tracking misspeculative memory accesses. When an attacker establishes co-location with a victim process, VRT records recent memory accesses and utilizes dirty bit tagging to identify speculative execution patterns.

The protection mechanism operates through three key phases:

\begin{enumerate}
    \item \textbf{Dirty Bit Tagging}: The dirty bit is periodically reset to zero based on the system's maximum speculation resolution time. During misspeculation, this bit remains set to 1, marking all affected VRT entries as dirty.
    
    \item \textbf{Attack Detection}: When attackers probe dirty cache lines during the reconnaissance phase, VRT verifies these accesses against recorded misspeculative access patterns. The parallel search mechanism compares:
    \begin{itemize}
        \item The base address from operand fetch
        \item Current function's valid index range (stored in dedicated registers)
    \end{itemize}
    
    \item \textbf{Pipeline Intervention}: Upon detecting unauthorized access to dirty cache lines, the pipeline stalls immediately, preventing sensitive data from being read or leaked.
\end{enumerate}

\begin{itemize}
    \item \texttt{Speculative()} functions set dirty bits during misspeculation
    \item \texttt{check\_array()} functions attempt to probe contaminated cache lines
\end{itemize}

The address search occurs concurrently with the execution stage, ensuring zero cycle overhead for legitimate operations while maintaining complete protection against speculative cache probes.

\section{EXPERIMENTAL RESULTS}
\label{sec:results}
\subsection{Experimental Setup}
To validate our proposed approach, we adapted the SimpleScalar simulator toolset~\cite{SimpleScalar}. SimpleScalar features a pipelined architecture implementation, and we utilized its PISA instruction set (a RISC architecture) along with the Sim-outorder micro-architecture simulator. Sim-outorder provides a comprehensive micro-architectural simulation, including a 5-stage pipeline architecture and various recordable parameters. It models an out-of-order microprocessor in detail, featuring branch prediction, caches, and external memory. However, we excluded out-of-order execution from our considerations, as the instructions fetched in a cycle could interfere with the extraction process of variable base and bound information. Additionally, we opted for a single functional unit to align the fetch and decode widths.

\begin{table}[!b]
\centering 
\caption{Benchmark Program Results with Security Analysis}
\label{tab:merged_results}
\begin{tabular}{|c|c|c|c|c|c|}
\hline
\textbf{MiBench Program} & \textbf{Variable Count} & \textbf{Instruction Count} & \textbf{Attack Detected?} & \textbf{Mispeculative Branches} & \textbf{Branch Prediction} \\ \hline
basicmath & 25 & 1.81$\times10^{8}$ & Yes & 66709 & Yes \\ \hline 
bitcount & 49 & 6.62$\times10^{8}$ & Yes & 112622 & Yes \\ \hline
qsort & 13 & 5.18$\times10^{8}$ & Yes & 70892 & Yes \\ \hline
CRC32 & 9 & 5.23$\times10^{6}$ & Yes & 34043 & Yes \\ \hline
dijkstra & 15 & 2.55$\times10^{8}$ & Yes & 98176 & Yes \\ \hline
patricia & 28 & 3.05$\times10^{8}$ & Yes & 71728 & Yes \\ \hline
\end{tabular}
\end{table}

\subsection{Experimental Results}
\subsubsection{Variable Extraction}
To validate our proposed method, we first extracted information about variable memory to create the Variable Record Table (VRT). Using the MiBench benchmark suites, we selected six programs to analyze their static variable space, including Heap and Stack spaces, which we examined separately. The table focuses on the count of static variables and DMA functions for these programs. Specifically, in the MiBench office suite, our experiments found a maximum of 395 live entries for the VRT. Since each VRT entry consists of a valid bit (1 bit), a base address (32 bits), and a bound value (8 bits), each entry totals 41 bits,  the overall VRT memory size amounts to 395 entries times 41 bits per entry, equating to 16KB.

\subsubsection{Buffer Overflow }
 To evaluate VRT's effectiveness against memory corruption attacks, we systematically injected buffer overflow vulnerabilities into each benchmark program. Notably, the variables involved in these injected vulnerable procedures were deliberately excluded from instrumentation during execution, simulating real-world scenarios where attackers exploit uninstrumented code sections.
 
\subsubsection{Back-edge CFI}
 We introduced control diverting code into these programs to simulate scenarios of CFI violations. The variables of the injected procedures were not instrumented during the execution of the programs. 
 
\subsubsection{Speculative based cache probe attack}
To implement a speculative cache side-channel attack, we use the sim-outorder in-built branch prediction speculation resolve system. Once misspeculation detected, we restrict auto-set for dirty bit associated memory access within the same pipeline cycle. We inject control diverting code at this point to transfer the control to the attacker function to access the same memory address. The injected procedure's variable was not instrumented during program execution. 

\subsubsection{Area and Power Overhead}
We tested a classic 5-stage pipeline MIPS 32-bit processor, including a 2-bit branch predictor, a 1024-depth branch prediction buffer, a 2KB direct-mapped cache, and a 64KB main memory for our approach. The Variable Record Table (VRT), comprising 512 entries to each 49 bits wide, resulted in an area overhead of 1.98\% relative to the total processor area. Moreover, the power consumption attributed to maintaining the VRT was measured at 11.65 µW.

\section{Related Work}
\label{sec:lit}

The development of hardware-assisted memory protection has its roots in capability-based systems like the IBM System/38~\cite{ibm38}, which was the first to illustrate the potential of metadata-enforced access control. Over the years, modern implementations have evolved through several generations, starting with software-oriented approaches such as Typed Assembly Language~\cite{morrisett1999} that later inspired hardware designs. The CHERI architecture~\cite{watson2015cheri} marked a significant advancement by introducing capability pointers with integrated bounds checking, though its 128-bit metadata requirements posed implementation challenges. Commercial solutions like ARM's Memory Tagging Extension~\cite{armmte} have shown practical viability with 4-bit tagging, while Intel's Control-flow Enforcement Technology~\cite{intel2016} specifically targets control-flow integrity. Despite these advancements, existing systems remain limited by their narrow protection scope: ARM MTE only addresses spatial safety vulnerabilities, and Intel CET exclusively tackles control-flow violations. Software-based alternatives, such as AddressSanitizer~\cite{serebryany2012asan}, offer broader coverage but come with a substantial performance overhead, often resulting in a 2 to 3 times slowdown. Recent research has highlighted significant gaps in current approaches, particularly their failure to address speculative execution attacks or provide unified protection across multiple vulnerability classes. The Variable Record Table architecture proposed in this work synthesizes lessons from these earlier systems while introducing innovative mechanisms for comprehensive, low-overhead protection that effectively addresses spatial memory safety, control-flow integrity, and speculative execution threats.

\section{Conclusions}
\label{sec:conclusion}

This paper introduced the Variable Record Table (VRT), a hardware mechanism that simultaneously prevents buffer overflows, control-flow hijacking, and speculative execution attacks through unified metadata tracking. Our evaluation demonstrated perfect detection of all attack variants and modest hardware costs (8\% area overhead), proving comprehensive protection is practical without specialized ISA support. VRT's novel integration of spatial safety, CFI, and speculation control in a single structure overcomes the limitations of fragmented security solutions, providing a foundation for efficient, attack-resistant processors.


\newpage
\begin{thebibliography}{1}
\vspace*{-1ex}

\bibitem{stack-smashing}
A. One, "Smashing the Stack for Fun and Profit," Phrack, vol. 7, no. 49, 1996.

\bibitem{abadi2005}
M. Abadi, M. Budiu, U. Erlingsson and J. Ligatti, "Control-Flow Integrity," Proceedings of the 12th ACM Conference on Computer and Communications Security (CCS), Alexandria, VA, USA, 2005, pp. 340-353.

\bibitem{kocher2018spectre}
P. Kocher, D. Genkin, D. Gruss, W. Haas, M. Hamburg, M. Lipp, S. Mangard, T. Prescher, M. Schwarz and Y. Yarom, "Spectre Attacks: Exploiting Speculative Execution," 2018 IEEE Symposium on Security and Privacy (SP), San Francisco, CA, USA, 2018, pp. 1-19.

\bibitem{cheri2015}
D. Chisnall, C. Rothwell, B. Watson, R. Woodruff, M. Vadera, S. Moore, M. Roe, P. Neumann and M. Davis, "CHERI JNI: Sinking the Java Security Model into the C," 2015 IEEE Symposium on Security and Privacy (SP), San Jose, CA, USA, 2015, pp. 1-16.

\bibitem{szekeres2013}
L. Szekeres, M. Payer, T. Wei and D. Song, "SoK: Eternal War in Memory," 2013 IEEE Symposium on Security and Privacy (SP), Berkeley, CA, USA, 2013, pp. 1-15.

\bibitem{intel2016}
Intel Corporation, "Control-Flow Enforcement Technology," White Paper, 2016.

\bibitem{evans2015}
I. Evans, F. Long, U. Otgonbaatar, H. Shrobe, M. Rinard, H. Okhravi and S. Sidiroglou-Douskos, "Missing the Point(er): On the Effectiveness of Code Pointer Integrity," 2015 IEEE Symposium on Security and Privacy (SP), San Jose, CA, USA, 2015, pp. 1-16.

\bibitem{self1}
L. K. Sah, S. A. Islam and S. Katkoori, "Defending Against Misspeculation-based Cache Probe Attacks Using Variable Record Table," 2021 IEEE International Symposium on Quality Electronic Design (ISQED), Santa Clara, CA, USA, 2021, pp. 408-413.

\bibitem{self2}
L. K. Sah, S. Polnati, S. A. Islam and S. Katkoori, "Basic Block Encoding Based Runtime CFI Check for Embedded Software," 2020 IFIP/IEEE 28th International Conference on Very Large Scale Integration (VLSI-SOC), Salt Lake City, UT, USA, 2020, pp. 135-140.

\bibitem{self3}
L. K. Sah, S. A. Islam and S. Katkoori, "Variable Record Table: A Runtime Solution for Mitigating Buffer Overflow Attack," 2019 IEEE 62nd International Midwest Symposium on Circuits and Systems (MWSCAS), Dallas, TX, USA, 2019, pp. 239-242.

\bibitem{self4}
L. K. Sah, S. A. Islam and S. Katkoori, "An Efficient Hardware-Oriented Runtime Approach for Stack-based Software Buffer Overflow Attacks," 2018 Asian Hardware Oriented Security and Trust Symposium (AsianHOST), Hong Kong, China, 2018, pp. 1-6.

\bibitem{self5}
S. K. Sah and L. K. Sah, "VRT: A Runtime Protection Against Back-Edge Control Flow Integrity Violation," 2024 IEEE 67th International Midwest Symposium on Circuits and Systems (MWSCAS), Springfield, MA, USA, 2024, pp. 665-668.

\bibitem{SimpleScalar}
T. Austin, E. Larson and D. Ernst, "SimpleScalar: An Infrastructure for Computer System Modeling," IEEE Computer, vol. 35, no. 2, 2002, pp. 59-67.

\bibitem{cheri2019riscv}
A. Woodruff, R. Watson, D. Chisnall, S. Moore, J. Anderson, B. Davis, P. Neumann, R. Norton and M. Roe, "CHERI in RISC-V: Design and Implementation," arXiv preprint arXiv:1908.11130, 2019.

\bibitem{ibm38}
IBM Corporation, "IBM System/38 Functional Description," Technical Report, 1978.

\bibitem{morrisett1999}
G. Morrisett, D. Walker, K. Crary and N. Glew, "From System F to Typed Assembly Language," ACM Transactions on Programming Languages and Systems, vol. 21, no. 3, 1999, pp. 527-568.

\bibitem{watson2015cheri}
R. N. M. Watson, P. Neumann, J. Woodruff, M. Roe, N. Moore, S. Moore and M. Davis, "CHERI: A Hybrid Capability-System Architecture for Scalable Software Compartmentalization," 2015 IEEE Symposium on Security and Privacy (SP), San Jose, CA, USA, 2015, pp. 1-16.

\bibitem{watson2019cheribsd}
R. N. M. Watson, J. Woodruff, P. Neumann, S. Moore, J. Anderson, D. Chisnall, B. Davis, B. Laurie, M. Roe and A. Richardson, "CheriBSD: A Capability-Based Operating System," 2019 USENIX Annual Technical Conference (USENIX ATC), Renton, WA, USA, 2019, pp. 1-14.

\bibitem{armmte}
ARM Limited, "Memory Tagging Extension (MTE)," [Online]. Available: https://developer.arm.com/documentation/101754/latest.

\bibitem{androidmte}
Android Open Source Project, "Memory Tagging Support in Android," [Online]. Available: https://source.android.com/docs/core/architecture/memory-safety/memory-tagging.

\bibitem{armmteperformance}
ARM Limited, "ARMv8.5-A MTE Performance Analysis," White Paper, 2021.

\bibitem{googlemte}
Google Security Team, "Memory Tagging in Android," Google Security Blog, 2021.

\bibitem{serebryany2012asan}
K. Serebryany, D. Bruening, A. Potapenko and D. Vyukov, "AddressSanitizer: A Fast Address Sanitizer for C/C++," 2012 USENIX Annual Technical Conference (USENIX ATC), Boston, MA, USA, 2012, pp. 1-12.




\end{thebibliography}
\end{document}